\documentclass{PoS}

\title{Nucleosynthesis of heavy elements in gamma ray bursts}

\ShortTitle{Nucleosynthesis in GRBs}

\author{\speaker{Agnieszka Janiuk} and Bartlomiej Kaminski\\
        Center for Theoretical Physics\\
        Polish Academy of Sciences\\
        Al. Lotnikow 32/46\\
        00-668 Warsaw, Poland\\
        E-mail: \email{agnes@cft.edu.pl}, \email{bkaminski@cft.edu.pl}}

\abstract{The ultrarelativistic jets responsible for prompt and afterglow emission in gamma ray bursts are presumably driven by a central engine that consists of a dense accretion disk around a spinning black hole. 
We consider such engine, composed of free nucleons, electron-positron pairs, Helium nuclei, and cooled by neutrino emission. A significant number density of neutrons in the disk provide conditions for neutron rich plasma in the outflows and jets. Heavy nuclei are also formed in the accretion flow, at the distances 150-250 gravitational radii from the black hole. We study the process of nucleosynthesis in the GRB engine, depending on its physical properties. Our results may have important observational implications for the jet deceleration process and heavy elements observed in the spectra of GRB afterglows.
          }

\FullConference{Swift: 10 Years of Discovery,\\
		2-5 December 2014\\
		La Sapienza University, Rome, Italy }

\begin{document}

\section{Introduction}

Gamma ray bursts observed by Swift satellite reach the 
distances up to redshift of about 8 (e.g. GRB 090423, z=8.3).
Many of the long duration GRBs, statistically more frequently detected than short ones, are 
observationally associated with supernovae (e.g. GRB 130427A).
This fact supports the idea that had been put forward more than 20 years ago (Narayan, Paczynski \& Piran, 1992)
about gamma ray bursts being the signatures of death throes of massive stars.
This progenitor star, while it is collapsing and forming a black hole in its
center, supports the production of ultrarelativistic jets, which are ultimately 
responsible for the observed transient gamma ray emission.
Due to the extremely large rate of accretion onto the central black hole, 
in this kind of engine, so called 'collapsar', we should encounter a rotationally supported transient disk 
with huge, nuclear densities and temperatures reaching $kT\sim 1 MeV$.
These conditions are therefore sufficient for production of heavy elements
in the body of accretion disk, whose mass is about several solar masses (Popham et al. 1999).

In the case of short duration GRBs, the situation is somewhat more difficult to probe, because of the poor statistics. Only about 20 of GRBs with durations less than 2 seconds, have the measurements of their spectroscopic redshifts, and have been localized in their host galaxies. Also, the mergers of two neutron stars, which are frequently invoked as progenitors of short GRBs, 
may result in the formation of a $\sim 2.5 -3 M_{\odot}$ black hole, 
surrounded by a disk with a mass of only 0.2-0.5 solar masses 
(Ruffert \& Janka 1998; Kluzniak \& Lee 1998). Thus, even though the 
accretion rate is even higher than in case of long GRBs and nuclear 
matter densities are reached, the amount of disctinct heavy 
element species might be too small to be detectable.

\section{Accretion disk in the GRB central engine}

\begin{figure}
\includegraphics[width=8cm]{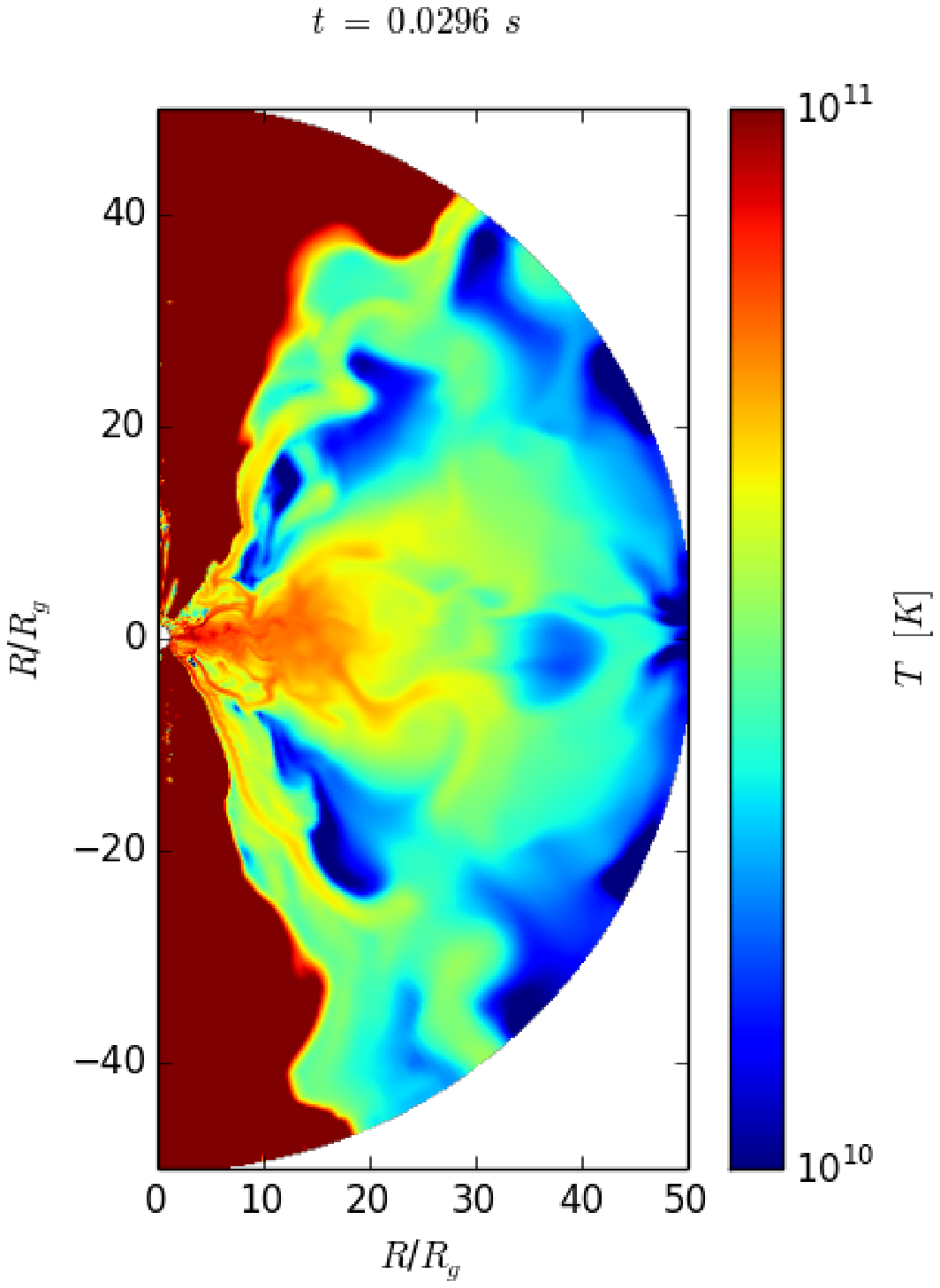}
\includegraphics[width=8cm]{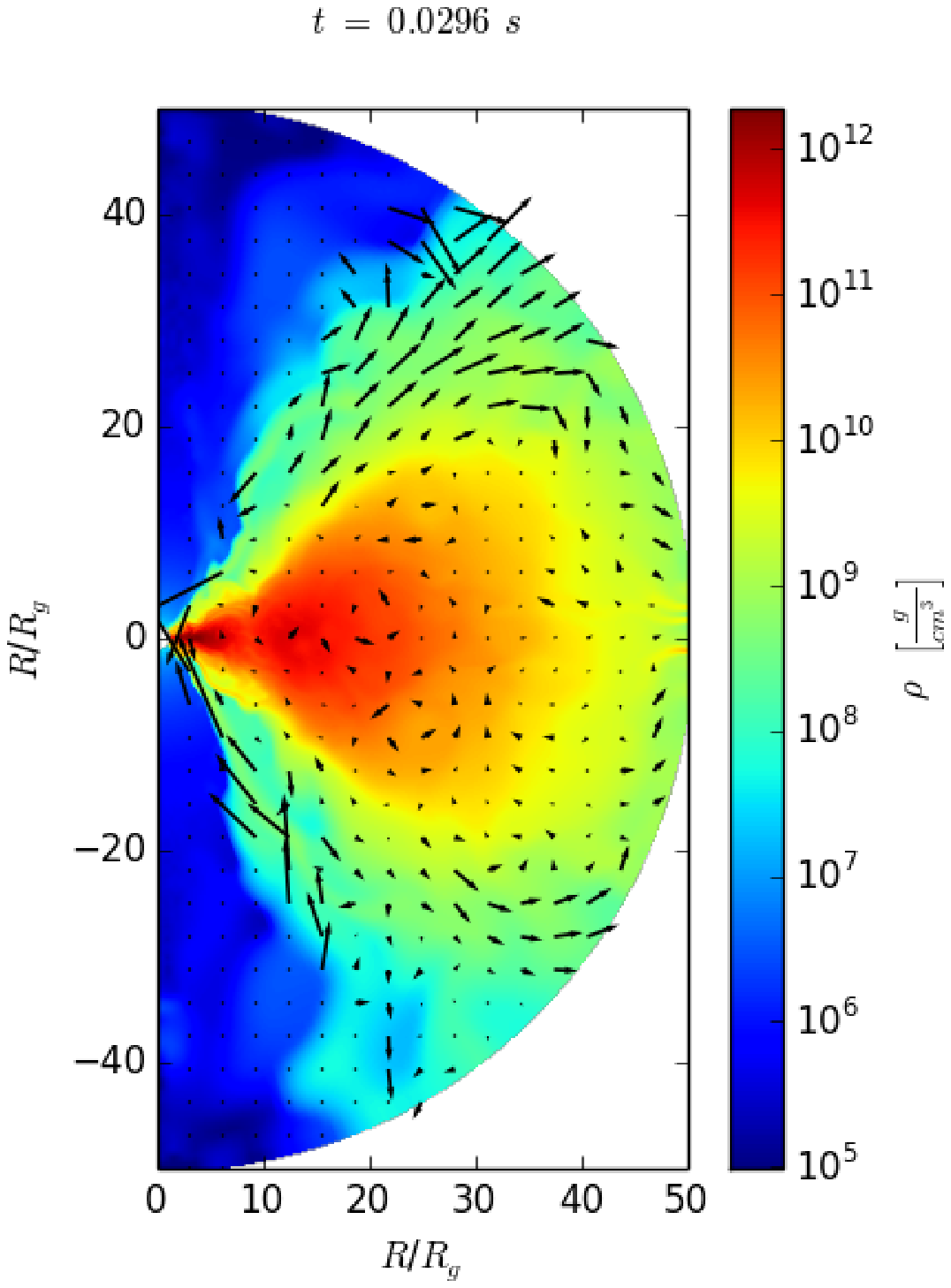}
\caption{Physical conditions in the torus accreting in the GRB central engine
and its outflows. The maps show temperature (left panel) and density with overplotted velocity field (right panel), in the innermost 50 $r_{g}$ around the black hole. Parameters: black hole mass $M=10 M_{\odot}$, spin $a=0.9$, 
mass of the disk $M_{d}=1.0  M_{\odot}$. Snapshot is taken from axisymmetric 
GR MHD simulation at time $t=0.0296$ s.
}
\label{fig:maps}
\end{figure}

The collapsar interpretation of long-duration gamma ray bursts invokes 
the existence of a rotating disk, or torus, in the interior of a collapsing 
star, fed by an external reservoir of stellar matter due to the 
fallback after supernova explosion. 
In the case of short 
duration GRBs, such torus should also have been formed, from the matter left by a 
drisrupted compact star, albeit the duration of the accretion phase is limited. 
In general, accretion disks in the context of gamma ray bursts are 
expected to have typical densities of $10^{10-12}$ g cm$^{-3}$, and 
temperatures that exceed $10^{11}$ K in the innermost $\sim 20$ Schwarzschild 
radii from the newly formed black hole. The accretion proceeds with rates up to
several Solar mass per second. In this hyper-accreting regime, the photons
are completely trapped in matter and are not efficient at cooling the disk.
However, in the weak interactions, neutrinos are produced and 
thus provide a mechanism for cooling the flow.
Moreover, at high accretion rates such torus becomes geometrically
'slim', with $H/r \sim 0.5$, and the advection of energy provides the 
cooling mechanism that is more efficient than neutrino emission.

To compute the structure of accretion disk, we need to 
take carefully into account the equation of state of its plasma.
In the physical conditions implied by hyper-accretion, the plasma equation of state will be very different from an ideal gas, and now also
the pressure of degenerate species, mainly electrons and positrons, as well 
as due to radiation and neutrinos, will contribute to EOS.
We determine the EOS of the disk numerically, including the
appropriate Fermi-Dirac statistics for the nuclear pressure of 
free neutrons, protons, and electron-positron pairs.
We also take into account the pressure of partially trapped neutrinos,
with the absorptive and scattering opacities included in a two-stream 
approximation (for details, see Janiuk et al. 2013, and references therein).

We studied the properties of accretion torus in GRB central engine with both
1-dimensional (vertically integrated), stationary, simplified models, 
using the assumption
of a classical Shakura-Sunyaev $\alpha$-viscosity prescription, as well 
as with the 2-dimensional, time-dependent GR MHD computations, where
the viscosity is provided by the magneto-rotational turbulence.
The results were presented already in a series of articles, and the recent
work (Janiuk et al. 2013) has shown that the resulting density and 
temperature distributions, as well as the neutrino emissivities, have very 
similar profiles, at least within the distance of the inner $\sim 50 R_{g}$ 
from the black hole. 

In Figure  \ref{fig:maps} we show the temperature and density
distributions in the $r-\theta$ plane of the innermost 50 gravitational radii 
in the GRB engine, 
as resulting from our axisymmetric MHD simulation. We also show the velocity field, to 
visualise the turbulent structure of the unbound wind outflows that are launched from this torus. The 
heavy elements will then be produced both in the accreting torus main body, 
and in the outflowing winds.

As was estimated by Surman \& McLaughlin (2004), the electron fraction in the outflowing material may be changed by 10-60\%, for an assumed constant velocity.
Later on, these authors (Surman et al. 2011) showed that production
of nickel depends sensitively on the entropy in the outflowing material. The outflows velocity was assumed there to have a simple power-law dependence
with distance, and a fixed starting position.
To determine accurately the mass fractions of heavy elements,
we need in fact to model both the chemical composition and evolution of the 
structure of the flow with magnetic tubulence, in order to accurately
describe the flow properties.

\section{Statistical Reaction Network}

To compute the abundances of heavy elements, we use the
thermonuclear reaction network code 
(http://webnucleo.org). The computational 
methods are described in detail in the literature (e.g., Wallerstein et al. 1997). 

In a nuthsell, the nuclear transmutations of subsequent species are tracked 
via the rate equations of the form:
\begin{equation}
\dot Y_{i} = \sum_{j} N_{j}^{i}\lambda_{j}Y_{j} + \sum_{j,k} N^{i}_{j,k}\rho \mathcal{N}_{A}<{j,k}>Y_{j}Y_{k} + \sum_{j,k,l} N^{i}_{j,k,l}\rho^{2} (\mathcal{N}_{A})^{2}<{j,k,l}>Y_{j}Y_{k}Y_{l}
\end{equation}
where $Y_{i}=n_{i}/\rho \mathcal{N}_{A}$ is the abundance of the $i$-th isotope, with $\mathcal{N}_{A}$ being the Avogadro number, $\lambda_{j}$ is the decay rate
of the $j$-th isotope, and the last two terms represent the encounters of 2 or 
3 reactant nuclei.
Such a formulation allows for a separation of the nuclear changes in plasma composition from the hydrodynamical effects. For the nuclei of atomic weight 
$A_{i}$, the condition $\sum A_{i}Y_{i}=1$ is satified. The charge conservation condition must also be satified, so $\sum Z_{i}Y_{i}=Y_{e}$ holds, under a given electron fraction $Y_{e}$.
The integrated cross-sections for interactions between target $j$ and projectile $k$ is expressed as:
\begin{equation}
<j,k>\equiv <\sigma v>_{j,k} = ({8 \over \mu \pi})^{1/2} (k_{B} T)^{-3/2} \int_{0}^{\infty} E \sigma(E) \exp(-E/k_{B}T) dE
\end{equation}
where $\mu$ is the reduced mass of the target-projectile system, $E$ is the center of mass energy, $T$ is the temperature, and $k_{B}$ is the Botlzmann constant. 

For nuclei, in general the Maxwell-Boltzmann statistics applies, while for the photodisintegration cross-section photons must obey the Planck statistics. Note, however, that since the photodisintegration process is endoenergetic, 
its rate can be small.

The data input necessary to study the astrophysical nucleosynthesis 
processes comes from experimental measurements and theoretical predictions.
Our code uses the {\it nuceq} library to compute the nuclear statistical equilibria
established for the thermonuclear fusion reactions.
The abundances are calculated under the constraints of nucleon number conservation
and charge neutrality, and the appropriate correction function to account for
degeneracy of relativistic species is used.
The reaction data were
 downloaded from JINA {\it reaclib} online database (http://www.jinaweb.org), 
prepared for studies of the nuclear masses and 
nuclear partition functions, and for computations of the 
nuclear statistical equilibria.

\section{Results}

\begin{figure}
\includegraphics[width=8cm]{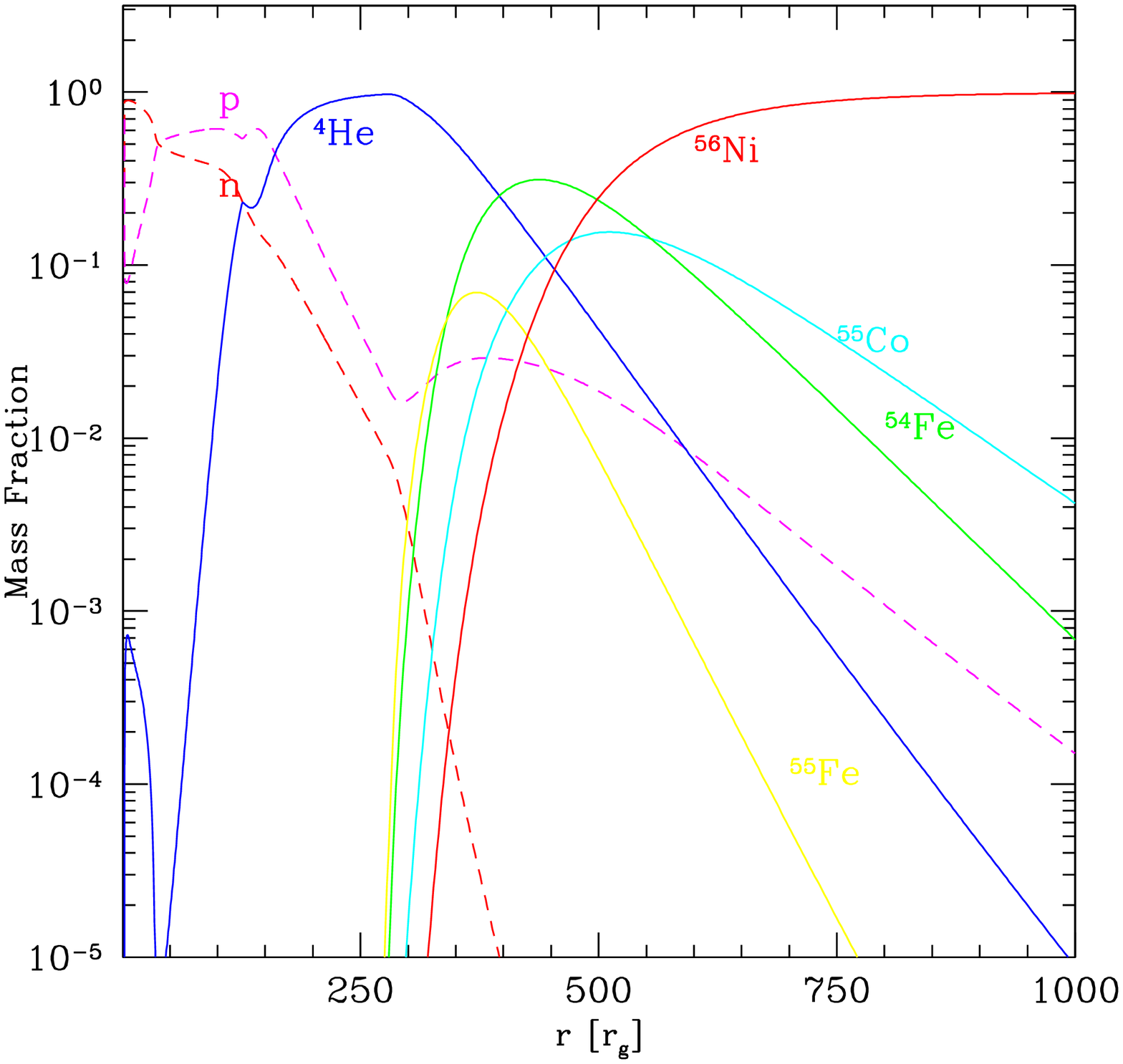}
\includegraphics[width=8cm]{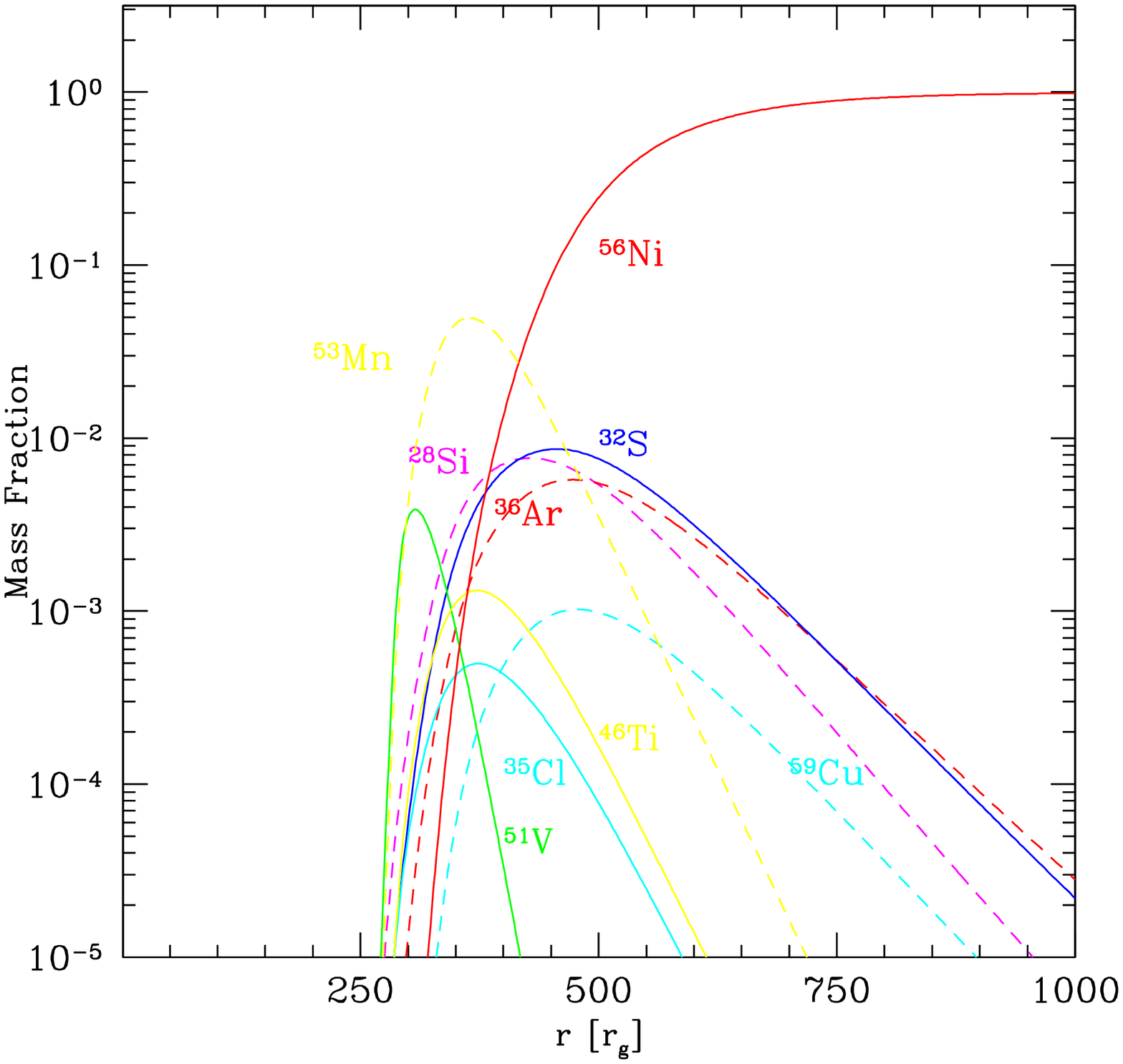}
\caption{The mass fractions of heavy elements synthesized in the 
accretion disk in GRB engine. The model parameters are: accretion rate of $\dot M= 0.1 M_{\odot}/s$ and
 black hole spin  $a=0.9$.
Left panel shows the abundance distribution of 
free protons and neutrons (dashed lines), as well as Helium, and the most abundant isotopes of Nickel, Iron, and Cobalt. 
Right panel shows the distribution of the most abundant isotopes of Silicon, 
Sulphur, Clorium, Argonium, Manganium, Titatnium, Vanadium, and Cuprum.
}
\label{fig:nseM21}
\end{figure}

The mass fraction of heavy nuclei was computed at every radius
of the accretion disk, given
the profiles of its density, temperature and electron fraction.
In Figure \ref{fig:nseM21}, we show the resulting distributions of the most abundant 
isotopes of heavy elements
synthesized in this plasma. An exemplary model is presented and its 
parameters are: accretion rate of $\dot M= 0.1 M_{\odot}/s$ and
 black hole spin  $a=0.9$.

For the accretion rate of 0.1 $M_{\odot}$s$^{-1}$, the abundance of $^{4}He$ is large with a value 
up to $\sim 260 r_{g}$ and then 
decreases
throughout the disk;
there is some fraction of $^{3}He$, Deuterium, and Tritium.
The next abundant isotopes are $^{28}Si$ - $^{30}Si$, $^{31}P$, $^{32}S$ - $^{34}S$, then
$^{35}Cl$, and $^{36}Ar$ - $^{38}Ar$.
Further, synthesized isotopes are $^{39}K$, $^{40}Ca$ - $^{42}Ca$, $^{44}Ti$ - $^{50}Ti$,
$^{47}V$ - $^{52}V$, $^{48}Cr$ - $^{54}Cr$, and $^{51}Mn$ - $^{56}Mn$.
The most abundant Iron isotopes formed in the disk are $^{52}Fe$ through  $^{58}Fe$; Cobalt is formed
with isotopes $^{54}Co$ through  $^{60}Co$, and Nickel isotopes are $^{56}Ni$ through  $^{62}Ni$.
The heaviest most abundant isotopes in our disk are  $^{59}Cu$ through  $^{63}Cu$.
Further, there is a smaller fraction of Zinc, $^{60}Zn$ - $^{64}Zn$, with a mass fraction above $10^{-5}$.
These heavy elements are generally produced outside 300-400 $r_{g}$. Inside this radius,
the disk consists of mainly free neutrons and protons with some fraction of Helium.
The mass fraction of free neutrons is smaller than that of protons, and
free neutrons disappear above $\sim 300 r_{g}$.

In comparison to the model with small accretion rate presented above,
the conditions in the disk with accretion rate of 1.0 $M_{\odot}$s$^{-1}$, are such that the mass fraction
of free neutrons
is larger than that of free protons inside $\sim 200 r_{g}$ and comparable
to a proton mass fraction up to $\sim 500 r_{g}$.
In both models, the heavy elements dominate above $\sim 550 r_{g}$ (see Janiuk 2014).

\section{Discussion}

The different properties
of central engines in the two classes of bursts, namely short and long GRBs, which we accounted for here
using two distinct values of accretion rate in their central engines,
determine qualitatively the details of the
nucleosynthesis process. This should be taken into account in the
 statistical studies of the observed phenomena  
(Gehlers et al. 2008).

For the accretion rate of $0.1 M_{\odot}/s$, our calculations show the significant proton excess in the disk
 above $\sim 250 r_{g}$. The wind ejected
at this region may therefore provide a substantial abundance of
light elements, Li, Be, and B. 
The high-accretion rate disk, on the other hand,
produces neutron rich outflows and forms heavy nuclei 
via the $r$-process. 
As we show here, 
the outflows ejected from the innermost $100 r_{g}$ in the high-accretion rate disks
 are also significantly neutron rich.
Therefore these neutron-loaded ejecta, which are accelerated via the black hole rotation, 
 feed the collimated jets at a large distance from the central engine.
This has important implications for the observed GRB afterglows, which are
induced by the radiation drag 
(Metzger et al. 2008),
and collisions between 
the proton-rich and neutron-rich shells within the GRB fireball 
(Beloborodov 2003).


The signatures of heavy elements synthesis have been found
in a number of supernovae associated with gamma ray bursts and their underlying spectra (e.g., Iwamoto et al. 1998, Kawai et al. 2006) or lightcurves (e.g., Nakamura et al. 2001). 
The isotopes synthesized in the GRB central engine accreting torus and 
its outflows during the prompt emission phase, 
should be detectable via the X-ray emission
that originates from their radioactive decay. These isotopes, such as Titanium 
$^{45}Ti$, $^{57}Co$, $^{58}Cu$, $^{62}Zn$, $^{65}Ga$, $^{60}Zn$, $^{49}Cr$, $^{65}Co$, $^{61}Co$, $^{61}Cu$, and $^{44}Ti$,  might give the
signal in the 12-80 keV energy band, which could be observed by current instruments
of good energetic resolution, e.g. by NuSTAR (e.g, Kouveliotou et al. 2013).

{\bf Acknowledgements}
We acknowledge partial support from grant DEC-2012/05/E/ST9/03914 awarded by the Polish National Science Center.

\end{document}